\documentclass[%
 reprint,
 amsmath,amssymb,
 aps,
]{revtex4-2}
\usepackage{braket}
\usepackage{graphicx}
\usepackage{subfig}
\usepackage{dcolumn}
\usepackage{bm}
\usepackage{comment}

\usepackage{algpseudocode}
\newcounter{algorithm}
\renewcommand{\thealgorithm}{\arabic{algorithm}}
\algrenewcommand\algorithmicrequire{\textbf{Input:}}
\algrenewcommand\algorithmicensure{\textbf{Output:}}

\usepackage{xcolor}

\makeatletter
\def\blfootnote{\xdef\@thefnmark{}\@footnotetext}
\makeatother

\begin{document}

\preprint{APS/123-QED}

\title{Qubit-centric Transformer for Surface Code Decoding}

\author{Seong-Joon Park$^{1,2,\dagger}$}
\blfootnote{$^\dagger$Work done while at POSTECH.}
\author{Hee-Youl Kwak$^{3, *}$}
\blfootnote{$^*$Corresponding author.}
\author{Yongjune Kim$^{4}$}\affiliation{%
 $^{1}$Samsung Electronics Company, Ltd., Suwon 16677, South Korea\\
 $^{2}$Institute of Artificial Intelligence, Pohang University of Science and Technology (POSTECH), Pohang 37673, South Korea\\
 $^{3}$Department of Electrical, Electronic and Computer Engineering, University of Ulsan, Ulsan 44610, South Korea\\
 $^{4}$Department of Electrical Engineering, Pohang University of Science and Technology (POSTECH), Pohang 37673, South Korea
}%

\date{\today}

\begin{abstract}
For reliable large-scale quantum computation, quantum error correction (QEC) is essential to protect logical information distributed across multiple physical qubits.
Taking advantage of recent advances in deep learning, neural network-based decoders have emerged as a promising approach to improve the reliability of QEC.
We propose the qubit-centric transformer (QCT), a novel QEC decoder for stabilizer codes based on a transformer architecture with a qubit-centric attention mechanism.
Our decoder transforms input syndromes from the stabilizer domain into qubit-centric tokens via a specialized embedding strategy.
These qubit-centric tokens are processed through attention layers to effectively identify the underlying logical error.
Furthermore, we introduce a graph-based masking method that incorporates the topological structure of quantum codes, enforcing attention toward relevant qubit interactions.
Across various code distances for surface codes, QCT achieves strong decoding performance, significantly outperforming existing neural decoders and the belief propagation (BP) with ordered statistics decoding (OSD) baseline.
Notably, QCT achieves a high threshold of $18.1$\% under the {code-capacity} noise model, which closely approaches the theoretical bound of $18.9\%$ and surpasses both the BP+OSD and the minimum-weight perfect matching (MWPM) thresholds, while also exhibiting a substantially higher error suppression factor $\Lambda$. Furthermore, we extend QCT to the circuit-level noise model by incorporating temporal convolution blocks to process repeated syndrome measurements. Under realistic circuit-level noise conditions, QCT demonstrates scalable error suppression across the evaluated code distances, consistently outperforming advanced decoders including the correlated MWPM and BP+OSD.
This qubit-centric approach provides a scalable and robust framework for surface code decoding, advancing the path toward fault-tolerant quantum computing.
\end{abstract}
\maketitle

\section{Introduction}
\label{sec_intro}
Quantum error correction (QEC) is fundamental to the realization of fault-tolerant quantum computing, as it preserves the integrity of logical information against environmental noise and operational imperfections~\cite{lidar2013quantum, nielsen2002quantum, shor1995scheme, steane1996error, gottesman1997stabilizer, terhal2015quantum,google2023suppressing}.
The QEC process consists of two main stages: encoding, which maps $k$ logical qubits into $n>k$ physical qubits, and decoding, which estimates errors from measured syndromes.
Consequently, the performance of the QEC decoder is a primary factor in determining the overall reliability of quantum computation.

Although classical error correction techniques are well-developed~\cite{gallager1962low, fossorier1999reduced, berrou1993near}, they cannot be applied directly to QEC.
Unlike classical bits, quantum states cannot be copied to generate redundancy due to the no-cloning theorem~\cite{nielsen2002quantum}.
Moreover, direct measurement of qubits collapses their quantum state, and qubits are also susceptible not only to bit-flip errors but also to phase-flip errors~\cite{nielsen2002quantum}.
These challenges necessitate a fundamentally different approach based on indirect stabilizer measurements.
This process produces syndromes, from which the underlying qubit errors are inferred without disturbing the encoded quantum information~\cite{gottesman1997stabilizer}.

Recent advances in neural networks have opened a new avenue for QEC decoding~\cite{torlai2017neural, krastanov2017deep, varsamopoulos2018decoding, baireuther2018machine, meinerz2022scalable,bausch2024learning,wang2023transformer}.
Neural network-based decoders are generally categorized into two types: low-level decoders, which reconstruct the physical error pattern of length $n$, and high-level decoders, which estimate the logical error pattern of length $k$.
For high-level decoders, the task is formulated as a classification problem over $4^k$ possible logical error classes.
This approach is particularly advantageous for codes with a small $k$, such as the surface code ($k=1$).
Since the output dimension remains fixed at $4^k$ regardless of the code distance $d$, these decoders provide excellent structural scalability without an exponential increase in output complexity \cite{jung2024convolutional}.

Furthermore, since neural network-based decoders process syndromes through a fixed model architecture, they offer a deterministic decoding latency~\cite{varsamopoulos2018decoding}.
This deterministic runtime contrasts with hybrid decoding architectures that employ a pre-decoder and a post-decoder, such as the belief propagation (BP) with ordered statistics decoding (OSD).~\cite{panteleev2021degenerate, roffe2020decoding,LaiKuo2021}.
In these hybrid approaches, the overall decoding time fluctuates depending on the input syndrome pattern, as the computationally intensive post-processing stage is triggered only when the initial pre-decoder fails to converge. More generally, standard graph-based decoders such as MWPM and Union-Find can also exhibit syndrome-dependent runtimes because the amount of graph processing depends on the observed detection-event pattern.
Predictable latency is a critical prerequisite for fault-tolerant quantum computing, as the global system clock is limited by the decoder's worst-case response time ~\cite{varsamopoulos2018decoding,terhal2015quantum}.
Given the remarkable success of neural networks in classification tasks \cite{krizhevsky2012imagenet, he2016deep}, they have emerged as a promising solution for high-level QEC decoders. Early implementations utilizing feed-forward neural networks (FFNNs) \cite{varsamopoulos2018decoding} have since evolved into more sophisticated architectures, such as convolutional neural networks (CNNs) \cite{jung2024convolutional}, to better exploit the spatial structure of quantum codes.

More recently, the transformer architecture has emerged as the universal backbone of modern deep learning, demonstrating state-of-the-art performance across various domains, including large language models, image classification, and object detection~\cite{brown2020language,dosovitskiy2020image,carion2020end}.
Leveraging this widespread success, transformer-based decoders~\cite{b_ECCT,b_Park2024} have recently been adapted for QEC.
For instance, Transformer-QEC~\cite{wang2023transformer} utilizes a vision transformer architecture to predict local errors, serving as a pre-decoding stage that reduces syndromes for a global decoder like minimum weight perfect matching (MWPM)~\cite{Edmonds1965}.
In addition, AlphaQubit utilizes a hybrid architecture combining CNNs and recurrent neural networks (RNNs) within a transformer framework, effectively capturing both the topological structure of surface codes and the temporal correlations inherent in repeated syndrome measurements~\cite{bausch2024learning}.

\begin{figure}[t]
    \includegraphics[width=\columnwidth]{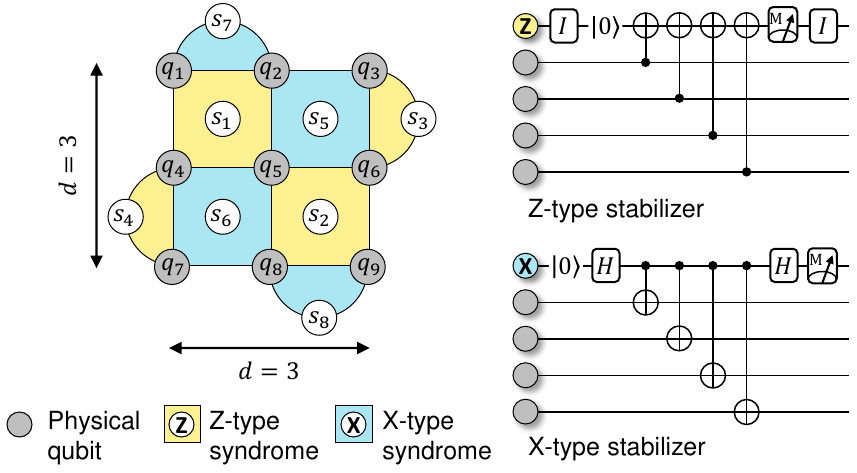}
    \caption{\label{fig_surface} 
    Illustration of the topological structure of the surface code with distance $d=3$ (left) and the syndrome extraction circuits (right).
    }
\end{figure}

However, a common limitation of these approaches is their stabilizer-centric input representation, which primarily operates on the syndrome space.
Although tracking the logical Pauli frame is sufficient for decoding in principle, logical errors ultimately arise from errors acting on the physical qubits of the code.
Therefore, organizing syndrome information around physical qubits more directly aligns the learned representation with the geometric structure of the QEC code.
Consequently, we investigate a qubit-centric perspective as a more direct representation for capturing the underlying error patterns, and we show that this qubit-centric representation consistently improves decoding performance.

In this work, we propose the qubit-centric transformer (QCT), an architecture that shifts the decoding perspective from stabilizers to physical qubits.
Unlike conventional stabilizer-centric approaches, QCT structures the input through two key layers: a qubit embedding layer and a merging layer.
The embedding layer projects localized syndrome vectors, which encapsulate the measurement outcomes of adjacent Z- and X-stabilizers, into dense feature tokens.
Subsequently, the merging layer concatenates these decoupled representations and fuses them into unified qubit tokens.
This design allows the transformer blocks to process tokens that directly correspond to the physical qubits where errors actually originate, demonstrating that a physical qubit-level perspective is fundamentally effective for decoding.
Furthermore, we introduce a structure-aware mask matrix that reflects the topological structure of QEC codes.
To ensure the model captures the physical structure, this matrix restricts the self-attention mechanism by masking the connections between tokens if their corresponding physical qubits do not share at least one common stabilizer.
By incorporating this local structure, QCT focuses on physically relevant neighboring qubits and effectively captures local syndrome dependencies induced by shared stabilizers.

To further address realistic circuit-level noise environments characterized by repeated syndrome measurements, we extend QCT to the circuit-level noise model. By preserving the core QCT structure to process spatial information independently within each round, we maintain the strict qubit-centric constraints. We then interleave temporal convolution blocks between the transformer blocks to learn temporal correlations across the syndrome-extraction rounds. This interleaved design enables QCT to capture temporal correlations arising from repeated syndrome extraction while preserving the qubit-centric spatial attention mechanism.

We evaluate the decoding performance of the proposed QCT on surface codes across various code distances.
Consequently, we demonstrate that QCT achieves substantially lower logical error rates~(LERs) and higher pseudo-thresholds compared to existing neural network decoders.
Notably, our analysis reveals that QCT achieves a high threshold of approximately $18.1\%$ under the {code-capacity} noise model, which closely approaches the theoretical bound of $18.9\%$~\cite{Bombin2012}.
This performance significantly outperforms the $17.00\%$ threshold of the BP+OSD baseline and the $14.7\%$ threshold of the conventional MWPM algorithm.
Furthermore, we extend QCT to realistic circuit-level noise environments with repeated syndrome extraction and demonstrate that QCT consistently achieves lower LERs than standard MWPM, correlated MWPM, and BP+OSD across the evaluated code distances and physical error rates.
In particular, its superior performance over correlated MWPM indicates that the proposed qubit-centric spatial processing and temporal convolution blocks effectively capture the spatial and temporal dependencies arising from circuit-level faults.
Finally, QCT exhibits strong distance-dependent logical error suppression,
achieving higher error suppression factors under both the {code-capacity} and circuit-level noise models than the corresponding baseline decoders.
These results validate the robustness and scalability of QCT and its potential for large-scale fault-tolerant quantum computing.

The rest of the paper is structured as follows. In Sec.~\ref{sec_pre}, we briefly review the fundamentals of stabilizer codes, surface codes, and the high-level decoding method.
Then, in Sec.~\ref{sec_QCT}, we present the proposed QCT architecture, including qubit-centric embedding, merging layer, structure-aware masking strategy, and its extension to the circuit-level noise model using temporal convolution blocks.
In Sec.~\ref{sec_results}, we evaluate the performance of our model, comparing its LER against several classical and neural network-based decoders, and present the circuit-level noise evaluation.
In Sec.~\ref{sec_discussion}, we analyze the architectural components of QCT, examine its model size and scalability with increasing code distance, and evaluate its decoding thresholds.
Finally, Sec.~\ref{sec_conclusion} concludes the paper.


\begin{figure*}[t]
    \includegraphics[width=.8\textwidth]{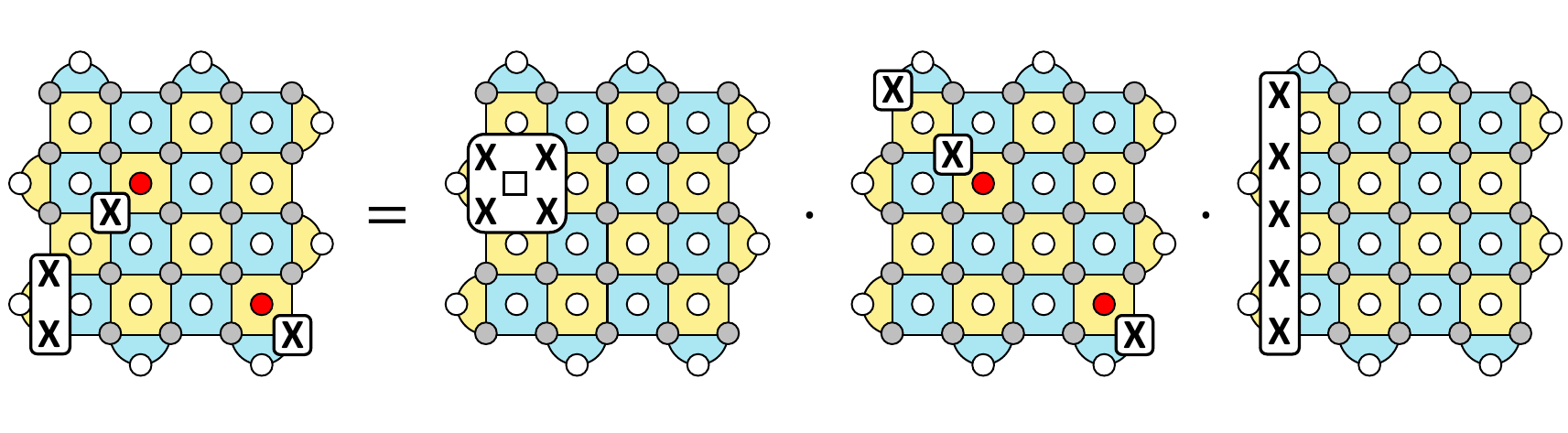}
    \caption{\label{fig_high_level}
    Illustration of the high-level decoding methodology treating decoding as a classification problem.
    An arbitrary physical error $E$ can be decomposed into $E = S \cdot T \cdot L$.
    }
\end{figure*}

\section{Preliminaries}
\label{sec_pre}
\subsection{Stabilizer code}
QEC codes utilize $n$ physical qubits to encode $k$ logical qubits. An $[[n, k, d]]$ stabilizer code $\mathcal{C}$ is defined by a stabilizer group $\mathcal{S}$, which is an Abelian subgroup of the $n$-qubit Pauli group $\mathcal{P}_{n}$ such that $-I_{n} \notin \mathcal{S}$ \cite{gottesman1997stabilizer}. The Pauli group $\mathcal{P}_{n}$ consists of $n$-fold tensor products of the single-qubit Pauli operators $\mathcal{P} = \{I, X, Z, Y\}$, defined as:$$I = \begin{pmatrix} 1 & 0 \\ 0 & 1 \end{pmatrix}, X = \begin{pmatrix} 0 & 1 \\ 1 & 0 \end{pmatrix}, Z = \begin{pmatrix} 1 & 0 \\ 0 & -1 \end{pmatrix}, Y = iXZ.$$The group $\mathcal{S}$ is generated by $n-k$ independent and commuting stabilizer generators $\{S_1, S_2, \dots, S_{n-k}\}$. The code space $\mathcal{C}$ is then defined as the common $+1$ eigenspace of all elements in $\mathcal{S}$:\begin{equation*}\mathcal{C} = \{ \ket{\psi} \mid S\ket{\psi} = \ket{\psi}, \forall S \in \mathcal{S} \}.\end{equation*}Any state $\ket{\psi} \in \mathcal{C}$ is referred to as a logical state, representing the collective state of the $k$ encoded logical qubits. By definition, stabilizers $S \in \mathcal{S}$ act trivially on all logical states, leaving the encoded information unchanged.

Logical operators are Pauli operators that commute with all elements in $\mathcal{S}$, thereby preserving the code space $\mathcal{C}$. 
Specifically, non-trivial logical operators are those contained in the centralizer of $\mathcal{S}$ but not in $\mathcal{S}$ itself, which can be defined as:
\begin{equation}
    \mathcal{L} = \{ L \mid [L, S] = 0, \forall S \in \mathcal{S} \} / \mathcal{S},
\end{equation}
where $[L, S] = LS - SL$ denotes the commutator of operators $L$ and $S$.
To distinguish them from physical operators, these operators acting on the encoded logical space are conventionally denoted as $L\in\{\bar{I}, \bar{X}, \bar{Z}, \bar{Y}\}$.
Here, the logical identity $\bar{I}$ acts trivially on the logical state (equivalent to the elements of $\mathcal{S}$), while the non-trivial logical operators $\bar{X}, \bar{Z},$ and $\bar{Y}$ act non-trivially on the logical qubits by transforming one logical state into another.
Since any stabilizer $S \in \mathcal{S}$ acts trivially on the code space, multiplying a logical operator $L$ by a stabilizer $S$ yields an equivalent logical operator (i.e., $(LS)\ket{\psi} = L(S\ket{\psi}) = L\ket{\psi}$).
The code distance $d$ is defined as the minimum weight of any such non-trivial logical operator.

When an error $E \in \mathcal{P}_{n}$ occurs on the physical qubits, it is identified by measuring the eigenvalues of the stabilizer generators.
This process yields a syndrome vector $\mathbf{s} = (s_1, \dots, s_{n-k}) \in \{0, 1\}^{n-k}$, where each syndrome bit $s_i$ is defined as:
\begin{equation}
    s_i = \begin{cases} 
    0 & \text{if } [S_i, E] = 0, \\
    1 & \text{if } [S_i, E] \neq 0.
    \end{cases}
\end{equation}

\subsection{Surface Codes}
The surface code is a class of topological stabilizer codes, widely recognized for its high error threshold and its leveraging of local connectivity between qubits \cite{preskill2018quantum, google2024quantum}. 
As illustrated in Fig.~\ref{fig_surface}, which shows a rotated surface code with distance $d=3$, physical qubits (often referred to as data qubits, indicated by grey circles) are arranged on the vertices of a $d \times d$ lattice to encode $k=1$ logical qubit ($n=d^2$). 
To detect physical errors, $m = n-k = d^2-1$ stabilizer generators are distributed across the faces (plaquettes) of the lattice. 
For each face, an ancilla qubit is positioned to measure a syndrome bit.
The right panel of Fig.~\ref{fig_surface} illustrates the syndrome extraction circuit for this measurement.
These stabilizers are categorized into Z-stabilizers (yellow plaquettes) and X-stabilizers (blue plaquettes), which are used to detect $X$ (bit-flip) and $Z$ (phase-flip) errors on the physical qubits, respectively.

The measurement outcomes of these stabilizers form the full syndrome vector $\mathbf{s} = (s_1, \dots, s_m) \in \{0, 1\}^m$. 
Following the indexing order shown in Fig.~\ref{fig_surface}—where degree-4 stabilizers are indexed prior to degree-2 stabilizers—the vector $\mathbf{s}$ is partitioned as $\mathbf{s} = [\mathbf{s}^Z, \mathbf{s}^X]$. 
Specifically, the first $m^Z$ bits $\mathbf{s}^Z = (s_1, \dots, s_{m^Z})$ correspond to the Z-stabilizers, and the subsequent $m^X$ bits $\mathbf{s}^X = (s_{m^Z+1}, \dots, s_m)$ correspond to the X-stabilizers ($m = m^Z + m^X$).

On this topological structure, logical operators are defined as non-trivial Pauli operators that span across the lattice. 
Specifically, the logical $\bar{Z}$ operator corresponds to a chain of physical $Z$ operators connecting the top and bottom boundaries, while the logical $\bar{X}$ operator corresponds to a chain of physical $X$ operators connecting the left and right boundaries. 
Together with the logical identity $\bar{I}$ and $\bar{Y} \equiv i\bar{X}\bar{Z}$, these constitute the fundamental logical operators of the surface code.

\subsection{High-level decoding method}

We adopt a high-level decoding method~\cite{varsamopoulos2018decoding, jung2024convolutional} based on a neural decoding architecture, which simplifies the learning task as shown in Fig.~\ref{fig_high_level}.
This approach exploits the decomposition of any physical error $E$ into three distinct components: $E = S \cdot T \cdot L$.
Here, $S$ is a stabilizer, $L$ is a logical operator, and $T$ is a pure error that can be deterministically inferred from the syndrome $\mathbf{s}$.
{In our code-capacity depolarizing noise model experiments (hereafter referred to as the code-capacity noise model), each physical qubit independently undergoes a depolarizing Pauli error, while syndrome measurements are assumed to be perfect.}
In this setting, the syndrome-to-pure-error mapping is implemented using a lookup table (LUT),
which provides a syndrome-consistent pure error $T$.
For a fixed $E$ and $T$, the stabilizer $S$ and the logical operator $L$ are uniquely determined, implying that each error $E$ can be mapped to a corresponding logical operator $L$.
For surface codes, since each code encodes exactly one logical qubit ($k=1$), the logical operator is strictly restricted to the four equivalence classes, i.e., $L \in \{\bar{I}, \bar{X}, \bar{Z}, \bar{Y}\}$.
Thus, the objective of the neural decoder $f_{\theta}$, where $\theta$ denotes the model parameters, is reduced to a four-class classification problem: predicting the most likely logical operator $\hat{L} = f_{\theta}(\mathbf{s})$ from the syndrome vector $\mathbf{s}$.
The final recovery operator is determined as $R = T \cdot \hat{L}$. 
If the prediction is correct ($\hat{L} = L$), the combined operation of error and recovery results in $RE = (T L)(S T L) = S$, which preserves the code state since $RE|\psi\rangle = S|\psi\rangle = |\psi\rangle$ for any $|\psi\rangle \in \mathcal{C}$.

In contrast to low-level decoding, which aims to identify the exact physical error $E$ among $4^n$ possible outcomes, this high-level decoding approach fixes the output dimension as $k$ regardless of the code size $n$.
Therefore, the model scales for length $n$ efficiently without a rapid increase in structural complexity or training difficulty.

 \begin{figure}[t]
    \includegraphics[width=\columnwidth]{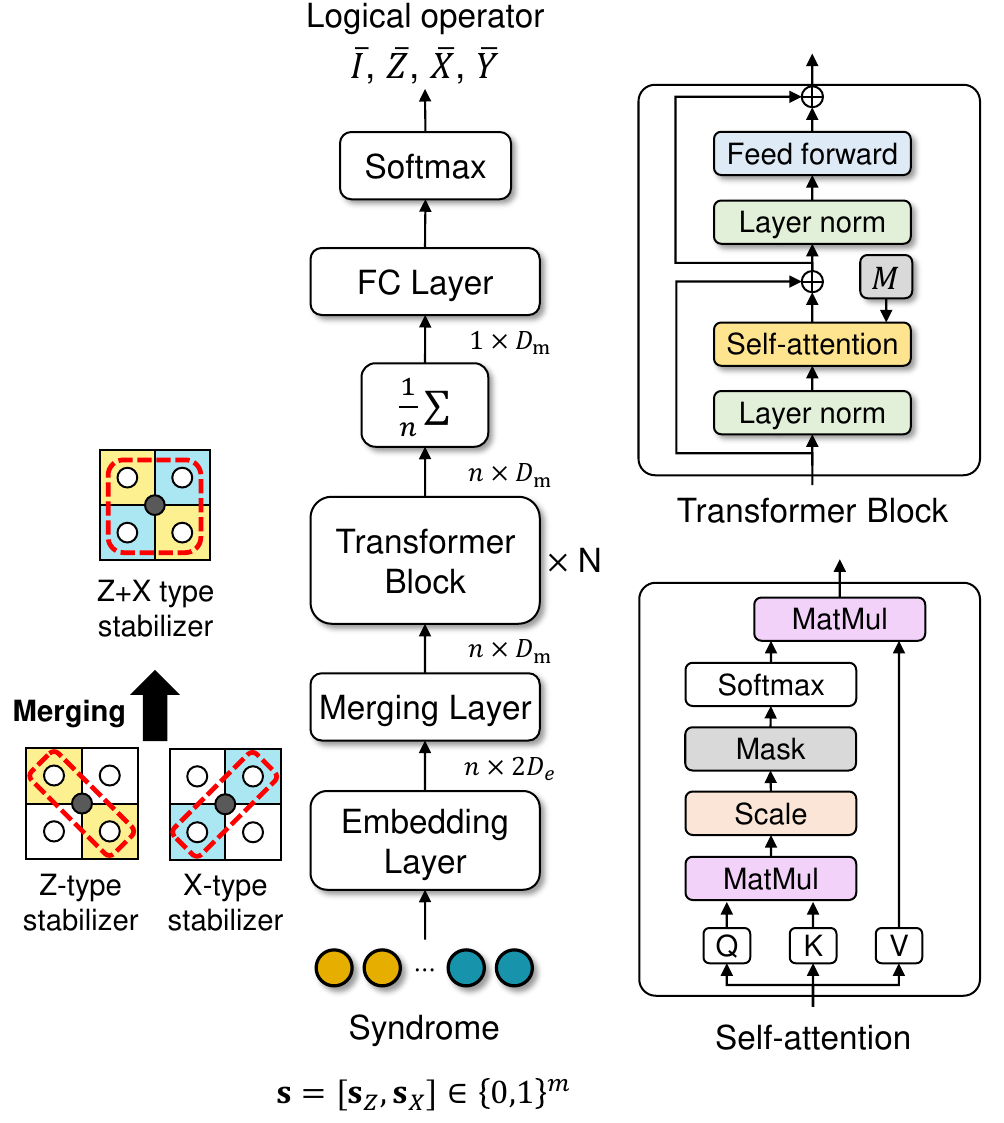}
    \caption{\label{fig_archi}
    The proposed QCT architecture.
    QCT processes the input syndrome vector $\mathbf{s}$ through the embedding layer and merging layer, followed by $N$ transformer blocks, to finally predict the logical operator.
    }
\end{figure}

\section{Qubit-centric Transformer}
\label{sec_QCT}

In this section, we introduce the QCT architecture, which consists of two key components: a \textit{qubit-centric embedding layer} and a \textit{merging layer}. 
These layers extract decoupled Z- and X-syndrome information for each physical qubit and subsequently fuse them into unified sequence tokens. 
These tokens are then processed through attention layers to learn the complex local syndrome dependencies induced by shared stabilizers. 
Furthermore, we incorporate a \textit{structure-aware mask matrix} that constrains the self-attention mechanism to reflect the topological structure of the QEC code. 
The overall architecture of the proposed QCT is illustrated in Fig.~\ref{fig_archi}.

\begin{figure}[t]
    \includegraphics[width=.9\columnwidth]{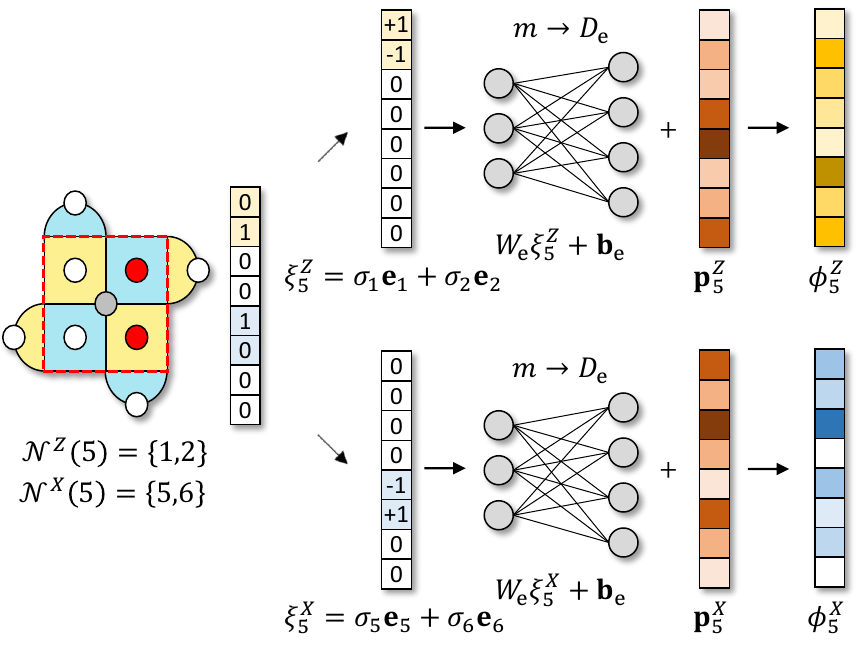}
    \caption{\label{fig_embedding}
    Illustration of the embedding process for physical qubit $q_5$ in a $d=3$ surface code.
    }
\end{figure}
    
\subsection{Qubit-centric Embedding Layer}

Let $\mathcal{N}^{Z}(i)$ and $\mathcal{N}^{X}(i)$ denote the sets of indices for Z- and X-stabilizers adjacent to the physical qubit $q_{i}$, respectively. 
For example, as illustrated in Fig.~\ref{fig_surface}, a central physical qubit $q_5$ is adjacent to two Z-stabilizers, $\mathcal{N}^Z(5) = \{1, 2\}$, and two X-stabilizers, $\mathcal{N}^X(5) = \{5, 6\}$. 
Due to the topological structure of the surface code, each physical qubit connects to at most two stabilizers of each type, i.e., $|\mathcal{N}^{Z}(i)|\le 2$ and $|\mathcal{N}^{X}(i)| \le 2$.

Let $\{\mathbf{e}_j\}_{j=1}^m$ be the standard basis vectors of $\mathbb{R}^m$ (e.g., $\mathbf{e}_1=[1,0,{\ldots},0]^\top$) and let $\sigma_j = (-1)^{s_j} \in \{+1, -1\}$ denote the value mapped from the $j$-th syndrome bit $s_j$. 
Throughout this paper, all vectors are defined as column vectors unless otherwise specified. 
For a stabilizer type $\tau \in \{Z, X\}$, the localized syndrome vector $\xi_i^{\tau}$ is defined as a sparse representation of the syndrome vector $\mathbf{s}$, localized to the neighbors of qubit $q_i$:
\begin{equation}
    \xi_i^{\tau} = \sum_{j \in \mathcal{N}^{\tau}(i)} \sigma_j \mathbf{e}_j.
\label{eq_xi}
\end{equation}
For each physical qubit $q_{i}$, we construct two localized syndrome vectors $\xi_i^{Z}$ and $\xi_i^{X}$ of size $m$ based on the equation~(\ref{eq_xi}), which encapsulate the measurement outcomes of the adjacent stabilizers. 
Note that the elements corresponding to non-adjacent stabilizers (i.e., $j \notin \mathcal{N}^{\tau}(i)$) are zero, ensuring that each qubit-centric vector carries only local syndrome information relevant to the specific physical qubit.

Fig.~\ref{fig_embedding} illustrates this embedding process for the physical qubit $q_5$ in a $d=3$ surface code. 
In this example, we assume that the syndrome bits $s_2$ and $s_5$ are 1, while all others are 0. 
Since $q_5$ is adjacent to Z-stabilizers indexed as $\mathcal{N}^Z(5)=\{1,2\}$ and X-stabilizers as $\mathcal{N}^X(5)=\{5,6\}$, localized syndrome vectors for $q_5$ are constructed as:
\begin{equation}
    \xi_5^{Z} = \sigma_1 \mathbf{e}_1 + \sigma_2 \mathbf{e}_2, \quad \xi_5^{X} = \sigma_5 \mathbf{e}_5 + \sigma_6 \mathbf{e}_6.
\end{equation}
Given the syndrome $s_2=1$ and $s_5=1$, we have $\sigma_2 = -1$ and $\sigma_5 = -1$, while the remaining coefficients are $\sigma_1 = +1$ and $\sigma_6 = +1$.
Note that $\xi_5^{\tau}$ is a sparse vector since most of its elements are zero.

Subsequently, each sparse localized syndrome vector is projected into a dense feature to extract high-level representations.
We first project $\xi_i^{Z}$ and $\xi_i^{X}$ into $D_{\text{e}}$-dimensional vectors via a fully connected (FC) layer and add learnable positional embedding vectors $\mathbf{p}_i^{Z},\mathbf{p}_i^{X}\in \mathbb{R}^{D_{\text{e}}}$, which are optimized during training.
Let $\phi_i^{Z}$ and $\phi_i^{X}$ denote the resulting feature embeddings for the Z- and X-type syndromes of the $i$-th qubit, respectively:
\begin{equation}
\label{eq_emb_NN}
    {\phi}_i^{\tau} = (W_{\text{e}} \xi_i^{\tau} + {\mathbf{b}_{\text{e}}})+\mathbf{p}_i^{\tau},
\end{equation}
where $W_{\text{e}} \in \mathbb{R}^{D_{\text{e}} \times m}$ and $\mathbf{b}_{\text{e}} \in \mathbb{R}^{D_{\text{e}}}$ are the weights and biases of the FC layer.
Finally, as shown in Fig.~\ref{fig_embedding}, the feature embeddings $\phi^Z_5$ and $\phi^X_5$ for the example qubit are obtained by applying this transformation to the constructed vectors $\xi_5^{Z}$ and $\xi_5^{X}$.

\begin{figure}[t]
    \includegraphics[width=\columnwidth]{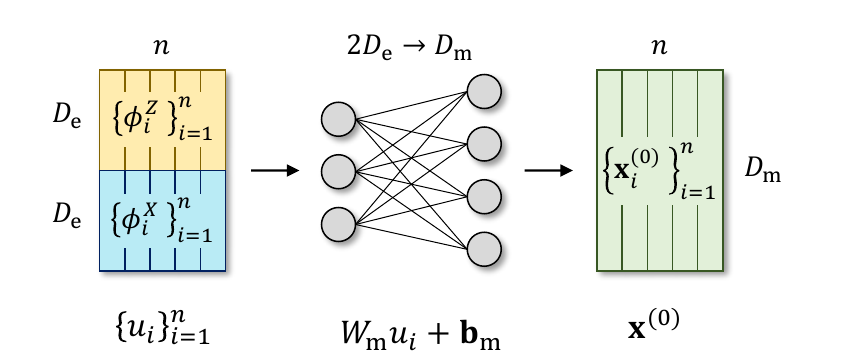}
    \caption{\label{fig_merging}
    Illustration of the merging layer.
    The input consists of $n$ decoupled Z-tokens $\{\phi^Z_i\}_{i=1}^n$ (yellow) and $n$ decoupled X-tokens $\{\phi^X_i\}_{i=1}^n$ (blue) separately.
    This process fuses the decoupled embeddings into unified qubit tokens $\mathbf{x}^{(0)}$.
    }
\end{figure}

\subsection{Merging Layer}
Finally, a \textit{merging layer} integrates these decoupled representations of $\{\phi^Z_i\}_{i=1}^n$ and $\{\phi^X_i\}_{i=1}^n$ into a unified form. 
For each qubit $q_{i}$, we first concatenate the corresponding Z- and X-type embeddings into paired vectors $\{u_i\}_{i=1}^n$, which are defined as:
\begin{equation}
    u_i = 
    \begin{bmatrix}
        \phi_i^{Z} \\
        \phi_i^{X}
    \end{bmatrix} \in \mathbb{R}^{2D_{\text{e}}},
\end{equation}
where the notation above denotes the vertical concatenation (stacking) of the two column vectors along the feature dimension.

Then, we fuse the decoupled representations by projecting $u_i$ into a unified token $\mathbf{x}_i^{(0)}$ of dimension $D_{\text{m}}$ via a simple merging layer:
\begin{equation}
    \mathbf{x}_i^{(0)} = W_{\text{m}} u_i + \mathbf{b}_{\text{m}},
\end{equation}
where $W_{\text{m}} \in \mathbb{R}^{D_{\text{m}} \times 2D_{\text{e}}}$ and $\mathbf{b}_{\text{m}} \in \mathbb{R}^{D_{\text{m}}}$ are trainable parameters.
The resulting qubit-centric tokens $\{\mathbf{x}_i^{(0)}\}_{i=1}^n$ serve as the initial input to the transformer.
Fig.~\ref{fig_merging} illustrates the merging process of QCT.

\subsection{Transformer and Output Layer}
\label{subsec_transformer}
These tokens are concatenated to form the input matrix
$\mathbf{x}^{(0)} = [\mathbf{x}_1^{(0)}, {\ldots}, \mathbf{x}_n^{(0)}] \in \mathbb{R}^{D_{\text{m}} \times n}$.
As shown in Fig.~\ref{fig_archi}, at each layer $l \in \{1, \ldots, N\}$, the transformer block generates the output sequence $\mathbf{x}^{(l)}$, which is used as the input to the next layer.
Each block consists of a self-attention module, a feed-forward neural network (FFNN), and normalization layers.

The self-attention mechanism is parameterized by three learnable projection matrices for the query, key, and value: ${W}_\text{q}\in \mathbb{R}^{D_{\text{m}}\times D_h}$, ${W}_\text{k}\in \mathbb{R}^{D_{\text{m}}\times D_h}$, and ${W}_\text{v}\in \mathbb{R}^{D_{\text{m}}\times D_h}$, respectively, where $D_h$ is the dimension of each head.
Given an input ${X}$, the projections are computed as \mbox{${Q}={X}{W}_\text{q}$}, \mbox{${K}={X}{W}_\text{k}$}, and \mbox{${V}={X}{W}_\text{v}$}. Finally, the self-attention output is obtained by the scaled dot-product attention:
\begin{equation}
    \text{Attention}({Q},{K},{V}) = \text{softmax}\left(\frac{{Q}{K}^{\top}}{\sqrt{D_h}}\right){V}.
\end{equation}
The transformer blocks capture the global dependencies and relationships between qubits via the multi-head self-attention mechanism.

Finally, the output tokens from this block are aggregated into a single vector via mean pooling and passed through a FC layer to obtain the final prediction.
For surface codes, since $k=1$, QCT outputs a 4-dimensional logit vector.
Applying a softmax function to these logits produces a probability distribution $\hat{P}_c$ for class $c$ over the four logical error classes $c\in\{\bar{I}, \bar{X}, \bar{Z}, \bar{Y}\}$.
The class with the highest probability is selected as the decoder’s prediction.

The entire QCT model is trained end-to-end by minimizing the standard cross-entropy loss, $\mathcal{L}_{CE}$, between the predicted and true logical operators. The loss is defined as:
\begin{equation}
    \mathcal{L}_{CE} = - \sum_{c \in \{\bar{I}, \bar{X}, \bar{Z}, \bar{Y}\}} y_c \log(\hat{P}_c),
\end{equation}
where $y_c$ is the one-hot encoded true label ($1$ if $c$ is the correct class, $0$ otherwise).

\begin{figure}[t]
    \includegraphics[width=.9\columnwidth]{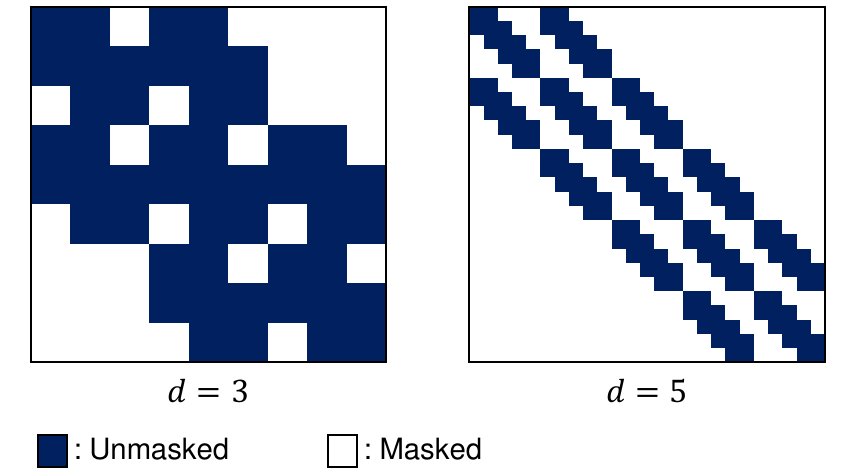}
    \caption{\label{fig_mask_matrix}
    Illustration of the structure-aware masks $M$ for surface codes used in QCT.
    The left figure shows the mask for $d=3$ surface code, and the right figure shows the mask for $d=5$ surface code.
    }
\end{figure}

\subsection{Structure-aware Mask Matrix}
\label{subsec_mask}
To learn the topological structure of the surface codes and improve learning efficiency, we propose a structure-aware mask for QCT.
In our architecture, each input token represents a unified physical qubit $q_{i}$, encapsulating its syndrome information.
Reflecting the structure of the surface code, the structure-aware mask $M$ is designed to allow any two qubits $q_{i}$ and $q_{j}$ to attend to each other if and only if they share at least one common stabilizer, regardless of whether it is a Z- or X-stabilizer.
All other qubit pairs that share no stabilizers are masked, enabling the model to effectively capture the local structure of the codes.

Let $\mathcal{N}(i)$ be the set of all stabilizers adjacent to qubit $q_{i}$, including all Z- and X-stabilizers adjacent to it:
\begin{equation}
    \mathcal{N}(i) = \mathcal{N}^{Z}(i) \cup \mathcal{N}^{X}(i),
\end{equation}
where $\mathcal{N}^{Z}(i)$ and $\mathcal{N}^{X}(i)$ denote the sets of adjacent Z- and X-stabilizers for $q_{i}$, respectively.
We then define $\mathcal{R}(i)$ as the index set of all qubits $q_{j}$ (including $q_{i}$ itself) that share at least one stabilizer with $q_{i}$:
\begin{equation}
    \mathcal{R}(i) = \{ j \mid \mathcal{N}(j) \cap \mathcal{N}(i) \neq \emptyset\}.
\end{equation}
Each element of the mask matrix $M \in \{0, -\infty\}^{n \times n}$ is then defined as:
\begin{equation}
    M_{ij} =
    \begin{cases}
        0 & \text{if } j \in \mathcal{R}(i), \\
        -\infty & \text{otherwise}.
    \end{cases}
\end{equation}
Applying this mask modifies the self-attention calculation as:
\begin{equation}
    \text{Attention}({Q},{K},{V}) =
    \text{softmax}\!\left(
        \frac{{Q}{K}^\top}{\sqrt{D_h}} + {M}
    \right){V}.
\end{equation}

Fig.~\ref{fig_mask_matrix} shows the structure-aware mask matrix for $d=3$ and $d=5$ surface codes.
White entries represent masked relations ($-\infty$) and blue entries represent unmasked relations ($0$).
This mask enforces the local connectivity of the surface code by allowing tokens to attend to each other only if their corresponding physical qubits share at least one common stabilizer (either Z-type or X-type).

\begin{table}[!t]
\refstepcounter{algorithm}
\label{alg:circuit_level}

\hrule
\vspace{1mm}

\noindent
\textbf{Algorithm~\thealgorithm.} QCT under the Circuit-level Noise Model

\vspace{1mm}
\hrule
\vspace{1mm}

\begin{algorithmic}[1]
\Require Cumulative syndromes $\mathbf{s}^{(1)}, \dots, \mathbf{s}^{(R)}$
\Ensure Pooled summary vector $\mathbf{z}$

\State $\mathbf{x}^{(0,t)} \gets \operatorname{Merging}\left(\operatorname{Embedding}(\mathbf{s}^{(t)})\right)$, $\forall t \in \{1,\dots,R\}$

\State $k_{\mathrm{temp}} \gets 1$

\For{$l = 1$ to $N$}

    \For{$t = 1$ to $R$}
        \State $\tilde{\mathbf{x}}^{(l,t)} \gets \operatorname{Transformer}_{l} \left(\mathbf{x}^{(l-1,t)}\right)$
    \EndFor

    \If{$l \bmod K = 0$ \textbf{or} $l=N$}

        \State $\delta \gets 1$ if $k_{\mathrm{temp}}$ is odd, otherwise $2$

        \For{$i = 1$ to $n$}

            \State $\mathbf{h}_i \gets \left[ \tilde{\mathbf{x}}_i^{(l,1)}, \dots, \tilde{\mathbf{x}}_i^{(l,R)} \right]$

            \State $\bar{\mathbf{h}}_i \gets \operatorname{LayerNorm}\left(\mathbf{h}_i\right)$

            \State $\mathbf{h}'_i \gets \mathbf{h}_i + \operatorname{Conv1D}_{\delta} \left( \operatorname{GELU} \left( \operatorname{Conv1D}_{\delta} \left(\bar{\mathbf{h}}_i\right) \right) \right)$

            \State Update $\tilde{\mathbf{x}}_i^{(l,t)}$ using $\mathbf{h}'_i$, $\forall t \in \{1,\dots,R\}$

        \EndFor

        \State $k_{\mathrm{temp}} \gets k_{\mathrm{temp}} + 1$

    \EndIf

    \State $\mathbf{x}^{(l,t)} \gets \tilde{\mathbf{x}}^{(l,t)}$, $\forall t \in \{1,\dots,R\}$

\EndFor

\State $\mathbf{x}^{(N,t)} \gets \operatorname{LayerNorm} \left(\mathbf{x}^{(N,t)}\right)$, $\forall t \in \{1,\dots,R\}$

\State $\displaystyle \mathbf{z} \gets \frac{1}{n} \sum_{i=1}^{n} \mathbf{x}_i^{(N,R)}$

\State \Return $\mathbf{z}$

\end{algorithmic}

\vspace{1mm}
\hrule

\end{table}

\subsection{QCT under the Circuit-Level Noise Model}

To address hardware environments characterized by circuit-level noise,
syndrome extraction is repeated over $R$ rounds.
Let $\mathbf{d}^{(t)} \in \{0,1\}^{m}$ denote the binary detector outcomes
at round $t \in \{1,\dots,R\}$.
For QCT under the circuit-level noise model, we use cumulative syndromes
over the repeated rounds as input.
The cumulative syndrome $\mathbf{s}^{(t)}$ at round $t$ is defined as
\begin{equation}
    \mathbf{s}^{(t)}
    =
    \bigoplus_{\tau=1}^{t} \mathbf{d}^{(\tau)},
\end{equation}
where $\oplus$ denotes element-wise modulo-2 addition.
The cumulative syndromes reflect faults accumulated during repeated
circuit-level syndrome extraction, including data-qubit, gate, reset, and
measurement faults.

For QCT under the circuit-level noise model, we preserve the core structure
of QCT to process spatial information independently within each round, while
employing temporal convolution blocks to learn temporal correlations across
the $R$ syndrome-extraction rounds.

For each round, we first map $\mathbf{s}^{(t)}$ into unified qubit tokens
$\mathbf{x}^{(0,t)}
=
[\mathbf{x}_1^{(0,t)},\dots,\mathbf{x}_n^{(0,t)}]$
for $t\in\{1,\dots,R\}$, employing the embedding and merging layers.
Then, the spatial dependencies within each round are captured independently
using the same transformer blocks and structure-aware mask matrix as in
the original QCT, as described in Sec.~\ref{sec_QCT}.C and \ref{sec_QCT}.D.
{Unlike QCT for the code-capacity noise model}, we insert a temporal convolution
block after every $K$ transformer blocks (in this work, $K=2$) to capture
temporal correlations across the $R$ syndrome-extraction rounds.

The temporal convolution block consists of layer normalization followed by
two 1D convolutional layers with kernel size $3$ and a residual connection.
For each physical qubit $q_i$, the feature vectors collected across the
$R$ rounds are arranged as
$\mathbf{h}_i \in \mathbb{R}^{D_{\mathrm{m}}\times R}$,
and the 1D convolutions are applied along the round dimension.
A Gaussian error linear unit~(GELU) activation is applied after the first
convolution, and the output of the second convolution is added to the input
through the residual connection.
The convolutional weights are shared across all physical qubit positions.
The dilation rate alternates between $\delta=1$ and $\delta=2$ across
successive temporal convolution blocks, enabling the network to capture
correlations over a wider range of rounds without increasing the number
of parameters.

Finally, after the last temporal convolution block, the feature
representations corresponding to the final extraction round $t=R$ are
selected and averaged over the physical qubit dimension.
The resulting representation is then processed by the same output layer as
in the original QCT.
The overall procedure is summarized in Algorithm~\ref{alg:circuit_level}.

\section{Performance Evaluation}
\label{sec_results}

For training, the QCT models are trained for 500 epochs, with 1000 minibatches per epoch and 128 samples per minibatch, using the Adam optimizer~\cite{kingma2014adam}. The learning rate is initialized at $10^{-4}$ and gradually decreases to $5 \times 10^{-7}$ using a cosine decay scheduler. {For the code-capacity noise model, the networks are trained} at a fixed physical error rate of $p=0.15$. For the circuit-level noise model, the models are trained on a mixture of physical error rates $p \in \{0.005, 0.006, 0.007, 0.008\}$. All simulations are performed on NVIDIA GeForce RTX A5000 GPUs and an AMD EPYC 7763 CPU.

For all Monte Carlo evaluations, the number of simulated shots is determined
adaptively according to a target number of observed logical errors.
Here, $N_{\mathrm{shots}}$ denotes the total number of simulated samples
used to estimate a logical failure probability, whereas $N_{\mathrm{err}}$
denotes the target number of logical-error events to be collected.
In our experiments, we set $N_{\mathrm{err}}$ between $500$ and $1000$
depending on the code distance and noise model.
For a measured logical failure probability $P_{\mathrm{fail}}$, the required
number of shots approximately satisfies
$N_{\mathrm{shots}} \geq {N_{\mathrm{err}}}/{P_{\mathrm{fail}}}$.
Thus, lower logical failure probabilities require a larger number of
simulated shots to collect the prescribed number of logical errors.
Error bars indicate 95\% Wilson confidence intervals.

\begin{figure*}[!t]
    \subfloat[]{\label{fig_LER_d5}\includegraphics[width=.33\textwidth]{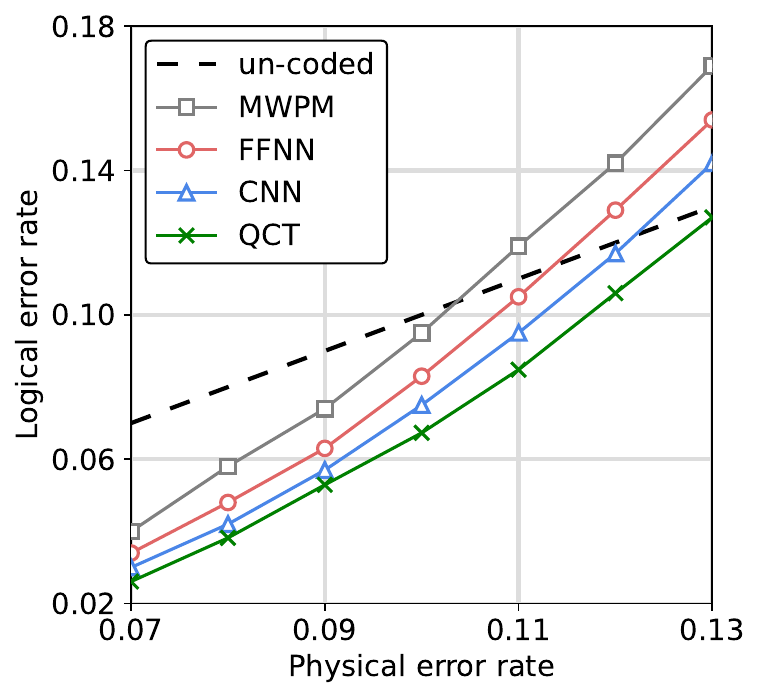}}
    \subfloat[]{\label{fig_LER_d7}\includegraphics[width=.33\textwidth]{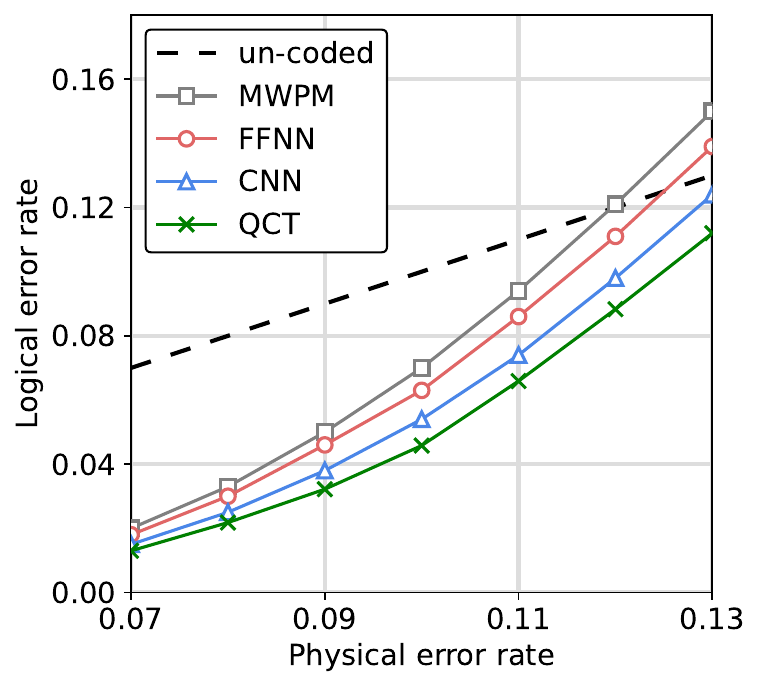}}
    \subfloat[]{\label{fig_LER_d9}\includegraphics[width=.33\textwidth]{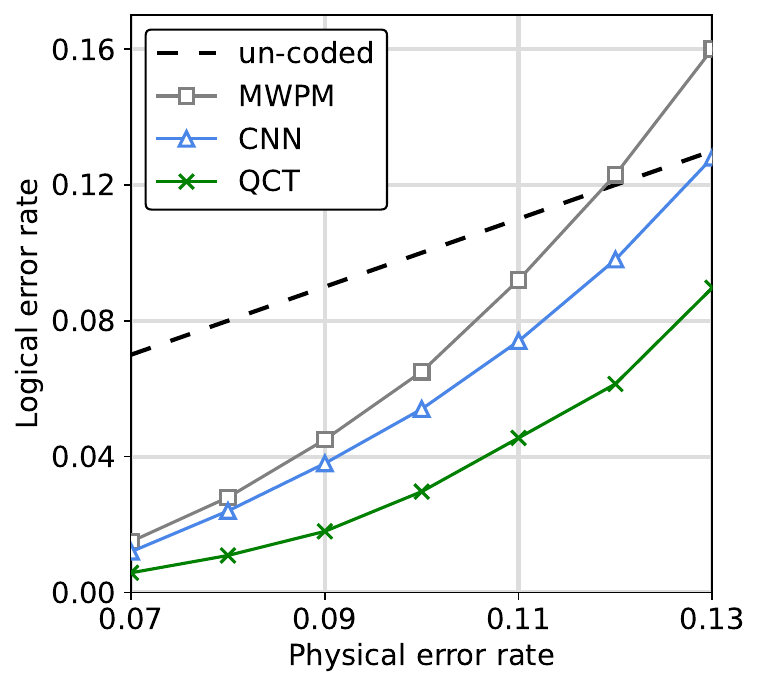}}
    \caption{\label{fig_ler_comparison} 
    {Performance under the code-capacity noise model}. Comparison of the proposed QCT decoder against MWPM, FFNN, and CNN decoders for code distances (a) $d=5$, (b) $d=7$, and (c) $d=9$.
    The MWPM, FFNN, and CNN results are taken from ~\cite{jung2024convolutional}.}
\end{figure*}

\begin{figure}[!t]
    \includegraphics[width=\columnwidth]{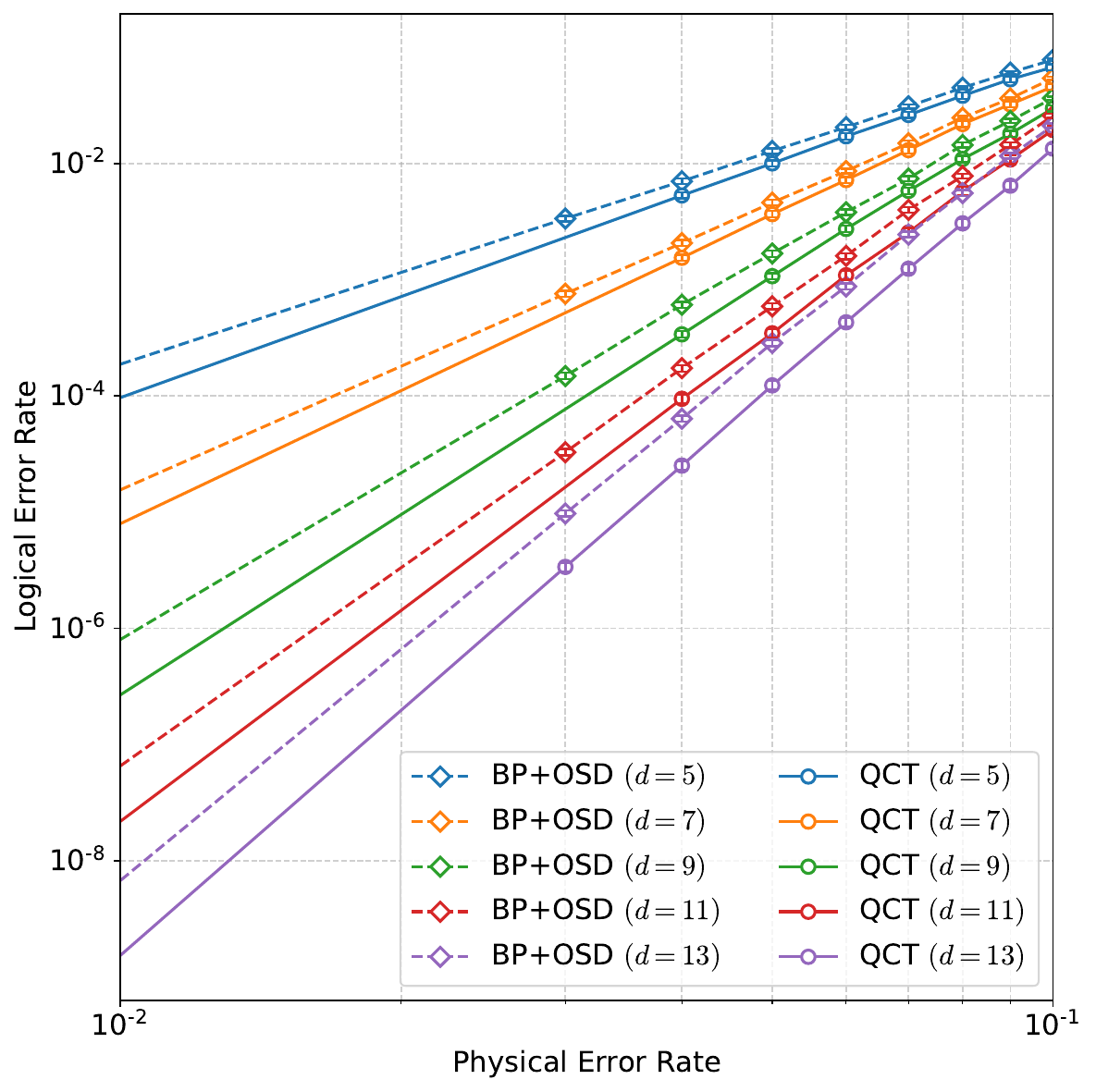}
    \caption{\label{fig_with_bp_osd}
    {Performance under the code-capacity noise model}. Comparison of the proposed QCT decoder against the BP+OSD baseline for code distances $d\in\{5,7,9,11,13\}$.
    At least $1000$ logical errors are collected for $d\in\{5,7,9\}$ and at least $500$ for $d\in\{11,13\}$. Error bars indicate 95\% Wilson confidence intervals. {The values at $p=0.01$ and $p=0.02$ are obtained by extrapolation and are shown without markers.}
    }
\end{figure}

\begin{figure}[!t]
    \includegraphics[width=\columnwidth]{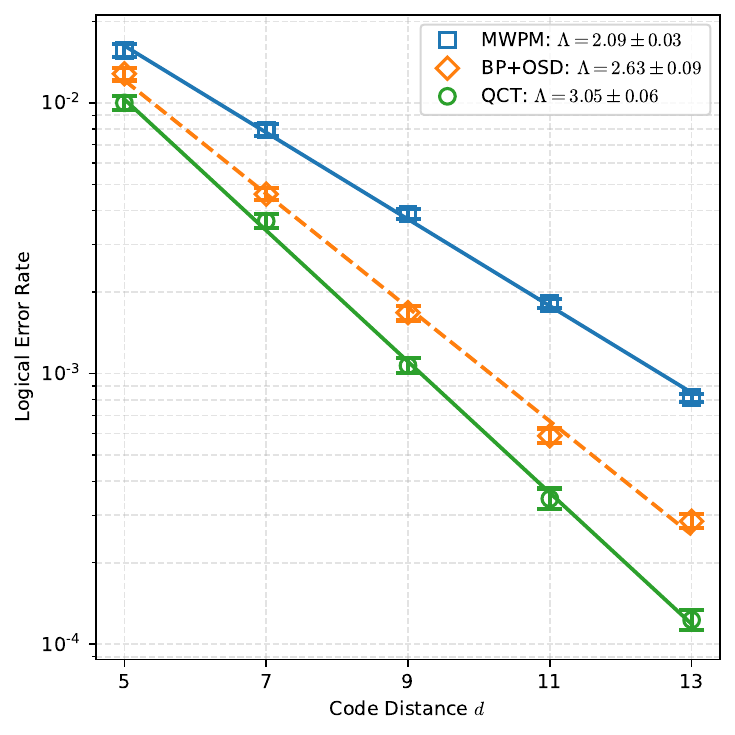}
    \caption{\label{fig_lambda_dep}
    {Error suppression factor $\Lambda$ under the code-capacity noise model},
    obtained from the LERs at $p=0.05$ for QCT, BP+OSD, and MWPM.
    At least $1000$ logical errors are collected for $d\in\{5,7,9\}$ and at least $500$ for $d\in\{11,13\}$. Error bars on the LER data indicate 95\% Wilson confidence intervals.
    }
\end{figure}

\subsection{Performance under {Code-capacity} Noise Model}
{We evaluate the LER performance of the proposed QCT for the surface codes under the code-capacity noise model.}
In all experiments, we choose transformer blocks with $N=8$ layers, $D_{\text{e}}=64$, and $D_{\text{m}}=128$.
{For testing under the code-capacity noise model}, we perform Monte Carlo simulations for physical error rates ranging from $p=0.03$ to $0.1$.
For lower physical error rates ($p=0.01$ and $0.02$), the LER values are estimated via curve fitting to the empirical data; therefore, these extrapolated points do not include error bars.
For distances $d \in \{5, 7, 9\}$ and $d \in \{11, 13\}$, we choose $N_{\mathrm{err}}=1000$ and $N_{\mathrm{err}}=500$, respectively.

Fig.~\ref{fig_ler_comparison} illustrates the decoding performance on surface codes with distances $d \in \{5, 7, 9\}$.
We first benchmark our model against the classical MWPM algorithm~\cite{dennis2002topological} and two representative neural network architectures: the FFNN~\cite{varsamopoulos2018decoding} and CNN~\cite{jung2024convolutional}.
Note that the total number of possible syndrome patterns the decoder should handle scales exponentially as $2^{d^2-1}$.
Despite this increasing complexity, QCT consistently outperforms the baseline decoders across all tested distances.
Notably, the performance gap between QCT and the other methods widens as the code distance increases.

\begin{figure}[!t]
    \includegraphics[width=\columnwidth]{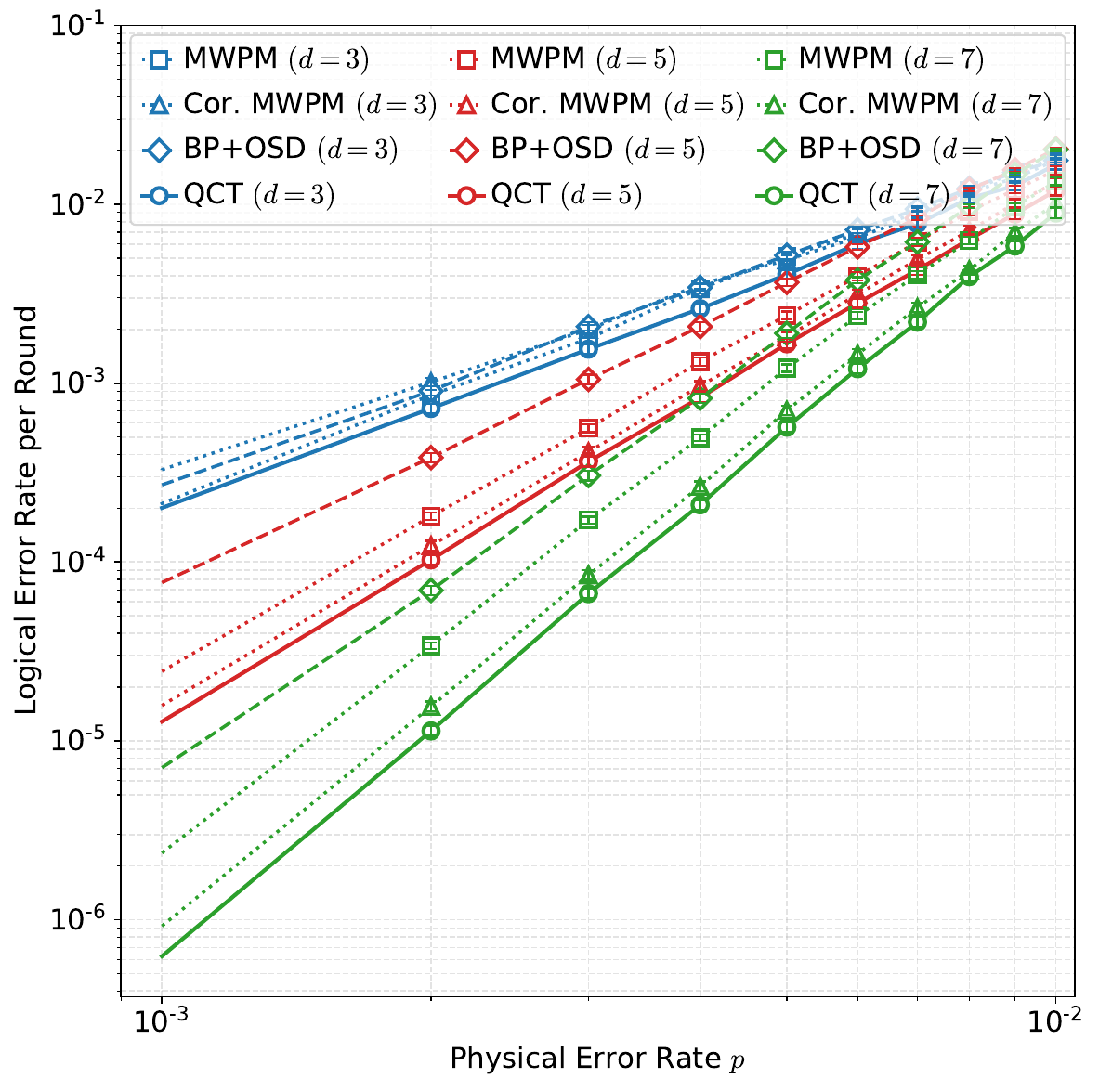}
    \caption{\label{fig_result_circuit}
    Performance under the circuit-level noise model in terms of the LER
    normalized per round. At least $1000$ logical errors are collected for each simulated data point, and error bars indicate 95\% Wilson confidence intervals. {The results at $p=0.001$ are obtained by extrapolation and are shown without markers}.
    }
\end{figure}

To further validate its performance, we compare QCT with the BP+OSD algorithm, a widely adopted baseline~\cite{roffe2020decoding, higgott2023improved, poulin2019neural, jung2025topological}.
For this comparison, we employ the quaternary BP~\cite{LaiKuo2021} with a maximum $20$ iterations.
As demonstrated in Fig.~\ref{fig_with_bp_osd}, QCT outperforms BP+OSD by a substantial margin across all code distances.
These results confirm that QCT achieves superior decoding performance, outperforming both advanced neural decoders and the BP+OSD baseline.

In Fig.~\ref{fig_lambda_dep}, we evaluate the error suppression factor~$\Lambda$.
At a fixed physical error rate of $p=0.05$, the LERs across code distances
are fitted according to
\begin{equation}
    P_L(d) = C\Lambda^{-(d+1)/2},
\end{equation}
where $C$ is a fitting constant and $\Lambda$ denotes the logical error
suppression factor~\cite{google2021exponential}.
A larger $\Lambda$ indicates stronger suppression of
the LER as the code distance increases.
As illustrated in Fig.~\ref{fig_lambda_dep}, QCT achieves an error
suppression factor of $\Lambda = 3.05 \pm 0.06$, higher than BP+OSD
($\Lambda = 2.63 \pm 0.09$) and MWPM ($\Lambda = 2.09 \pm 0.03$).
This result confirms that QCT exhibits stronger logical error suppression with increasing code distance compared to the baseline decoders.

\begin{figure}[!t]
    \includegraphics[width=\columnwidth]{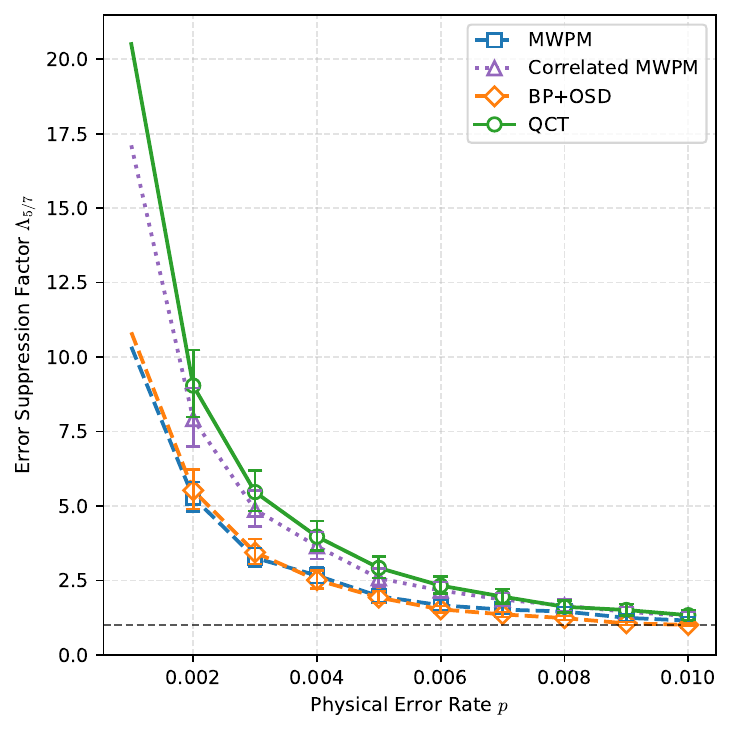}
    \caption{\label{fig_lambda_circuit}
    Error suppression factor $\Lambda_{5/7}$ under the circuit-level noise
    model for QCT, standard MWPM, BP+OSD, and correlated MWPM. At least $1000$ logical errors are collected for each simulated LER used to compute $\Lambda_{5/7}$; {the values at $p=0.001$ are based on extrapolated LERs and are shown without markers.}
    }
\end{figure}

\subsection{Performance under Circuit-Level Noise Model}

To evaluate the robustness of QCT under more realistic hardware conditions,
we benchmark its performance under a circuit-level noise model using the
Stim simulator~\cite{gidney2021stim}.
Following the detector-observable formulation used in Stim, the
circuit-level QCT directly predicts whether the logical observable is
flipped from the repeated syndrome information, rather than using the
{LUT-assisted high-level decoding employed under the code-capacity noise model.}
We use a uniform circuit-level depolarizing parameter $p$ in Stim,
with reset and measurement errors set to $2p/3$, corresponding to the
probability that a single-qubit depolarizing error flips the respective
preparation or measurement outcome.

For this circuit-level evaluation, we perform Monte Carlo simulations for
physical error rates ranging from $p=0.01$ down to $0.002$.
At each simulated data point, sampling is continued until at least $1000$
logical errors are observed.
For $p=0.001$, the LER is estimated by fitting the simulated data and is
therefore reported without an error bar.

\begin{figure*}[!t]
    \subfloat[]{\label{fig_N_values}\includegraphics[width=.33\textwidth]{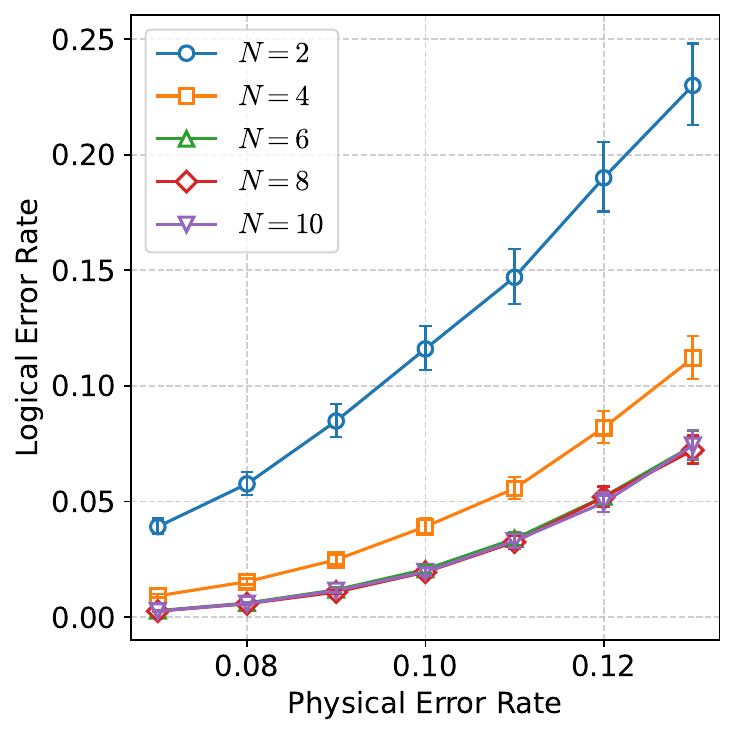}}
    \subfloat[]{\label{fig_mask_comp}\includegraphics[width=.33\textwidth]{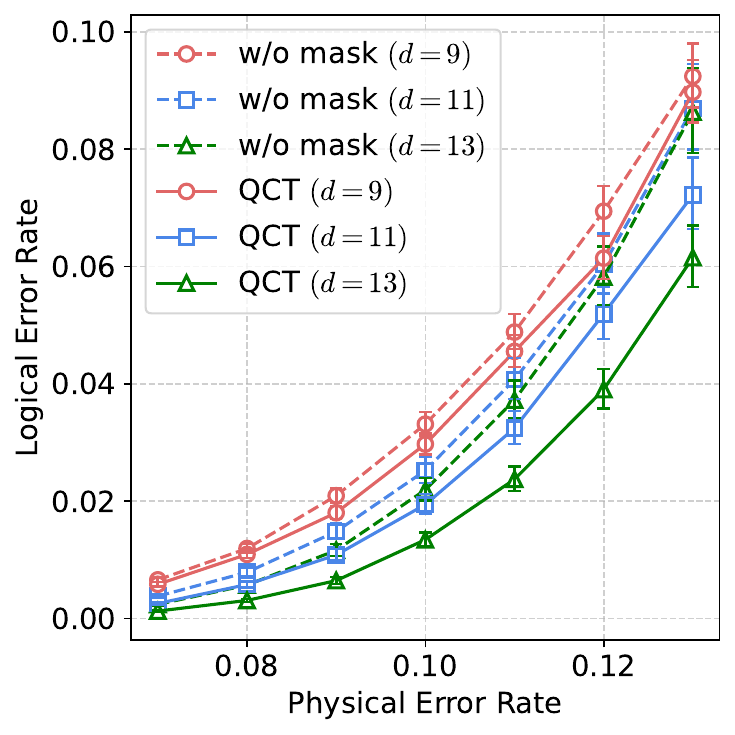}}
    \subfloat[]{\label{fig_merg_mask}\includegraphics[width=.33\textwidth]{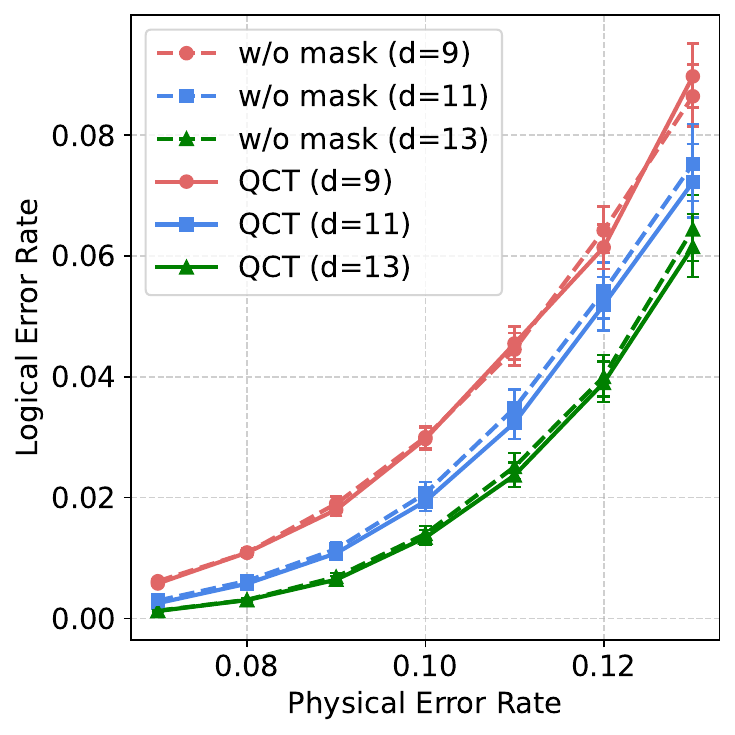}}
    \caption{\label{fig_comp}
    Decoding performance of QCT under various configurations: (a) impact of the number of transformer blocks ($N$), (b) effect of structure-aware masking, and (c) effect of the merging layer to decoding accuracy.
    At least $500$ logical errors are collected for each simulated data point, and error bars indicate 95\% Wilson confidence intervals.
    }
\end{figure*}

Fig.~\ref{fig_result_circuit} illustrates the LER normalized per round of QCT, evaluated on surface codes with distances $d\in\{3,5,7\}$.
The proposed QCT architecture is benchmarked against the standard MWPM, quaternary BP+OSD, and correlated MWPM decoders~\cite{fowler2013correlated}.
Both MWPM and correlated MWPM decoders are implemented using the PyMatching library~\cite{higgott2023pymatching}.
The results demonstrate that QCT consistently achieves lower LER per round compared to all three baseline decoders across all evaluated physical error rates.
Notably, QCT outperforms the correlated MWPM, which explicitly accounts for correlated fault mechanisms arising from the circuit-level noise model, indicating that QCT with qubit-centric spatial processing and temporal convolution blocks effectively captures the spatial and temporal dependencies arising from circuit-level faults.

Furthermore, to evaluate distance-dependent logical error suppression under the circuit-level noise model, we compute the error suppression factor between $d=5$ and $d=7$ as
\begin{equation}
    \Lambda_{5/7}
    =
    \frac{P_L(d=5)}{P_L(d=7)}.
\end{equation}
The resulting $\Lambda_{5/7}$ values are shown in Fig.~\ref{fig_lambda_circuit}.
QCT consistently exhibits a higher $\Lambda_{5/7}$ than the standard MWPM, quaternary BP+OSD and correlated MWPM, demonstrating stronger distance-dependent error suppression even under the circuit-level model.

\section{Discussion}
\label{sec_discussion}

\subsection{Analysis of Model Architecture}
In this subsection, we investigate the impact of various architectural components--the number of transformer blocks, the structure-aware mask, and the merging layer--on the decoding performance of QCT for a $d=11$ surface code.

\noindent \textbf{Number of Transformer Blocks:}
We examine the decoding performance according to the number of transformer blocks $N$.
As shown in Fig.~\ref{fig_comp}(a), increasing the model depth from $N=2$ to $N=8$ yields significant improvements in the LER; however, the performance gains saturate beyond $N=8$.
The substantial performance gap according to $N$ demonstrates that a shallow architecture with only $N=2$ blocks is insufficient to estimate the logical error present in the syndrome.
Based on this observation, we select $N=8$ as the default configuration, as it provides a balanced trade-off between performance and efficiency.

\noindent \textbf{Structure-aware Masking:}
To validate the effectiveness of the structure-aware mask, we conduct an ablation study comparing the proposed QCT with an unmasked variant where all tokens globally attend to each other.
Fig.~\ref{fig_comp}(b) compares QCT models with and without structure-aware masking.
The model without masking (labeled ``w/o mask'') successfully learns the decoding task, indicating that the self-attention mechanism can implicitly capture local syndrome dependencies induced by shared stabilizers.
However, the model with structure-aware masking consistently outperforms the model without masking across the evaluated physical error rates.
This result highlights that while the standard transformer successfully learns the error patterns via global attention, explicitly concentrating on physically relevant neighbors is a significantly more effective strategy.
The mask enables the model to focus on the topological structure of the codes, thereby enhancing the decoding performance.

\noindent \textbf{Merging Layer:}
To validate the effectiveness of the merging layer, we compare QCT with the model without the merging layer (labeled ``w/o merg'').
The only difference is the existence of the FC layer that projects the concatenated $2D_{\text{e}}$-dimensional vectors into $D_{\text{m}}$-dimensional unified tokens.
Fig.~\ref{fig_comp}(c) illustrates the decoding performance of the models with and without the merging layer.

As shown in the results, an additional merging layer provides a consistent improvement in decoding performance.
Considering that this layer introduces a small number of parameters relative to the overall QCT architecture, achieving such a reduction in the LER is efficient. 
In quantum error correction, even these marginal gains are crucial for approaching the fault-tolerance threshold.
Furthermore, it is worth noting that even the model without the merging layer still maintains excellent performance and outperforms the other baseline decoders.
This demonstrates that shifting the decoding perspective directly to the qubit-centric approach is fundamentally effective.

\begin{table}[t]
\caption{\label{tab_params}
Number of trainable parameters for the evaluated surface-code distances.
}
\begin{ruledtabular}
\begin{tabular}{c c c c c c c}
 $d$ & $3$ & $5$ & $7$ & $9$ & $11$ & $13$ \\
\colrule
{Code-capacity}
& -- & 1.61~M & 1.61~M & 1.62~M & 1.63~M & 1.64~M \\
Circuit-level
& 2.00~M & 2.02~M & 2.05~M & -- & -- & -- \\
\end{tabular}
\end{ruledtabular}
\end{table}

\begin{table}[t!]
    \vspace{5mm}
    \caption{\label{tab_pseudo_th}
    Comparison of the pseudo-thresholds for the proposed QCT against benchmark decoders (MWPM, FFNN, and CNN) on the surface code for distances $d=3, 5,$ and $7$ {under the code-capacity noise model.}
    The results of MWPM, FFNN, CNN are from \cite{jung2024convolutional}.
    }
    \begin{ruledtabular}
        \begin{tabular}{cccc}
        $d$ & $3$ & $5$ & $7$\\
        \colrule
        MWPM~\cite{dennis2002topological} & 8.28\% & 10.36\% & 11.94\%\\
        FFNN~\cite{varsamopoulos2018decoding} & 9.77\% & 11.35\% & 12.49\%\\
        CNN~\cite{jung2024convolutional} & 9.80\% & 12.15\% & 13.26\%\\
        QCT (Proposed) & 9.80\% & 13.27\% & 14.31\%\\
        \end{tabular}
    \end{ruledtabular}
\end{table}

\begin{figure}[!t]
    \subfloat[]{\label{fig_th}\includegraphics[width=\columnwidth]{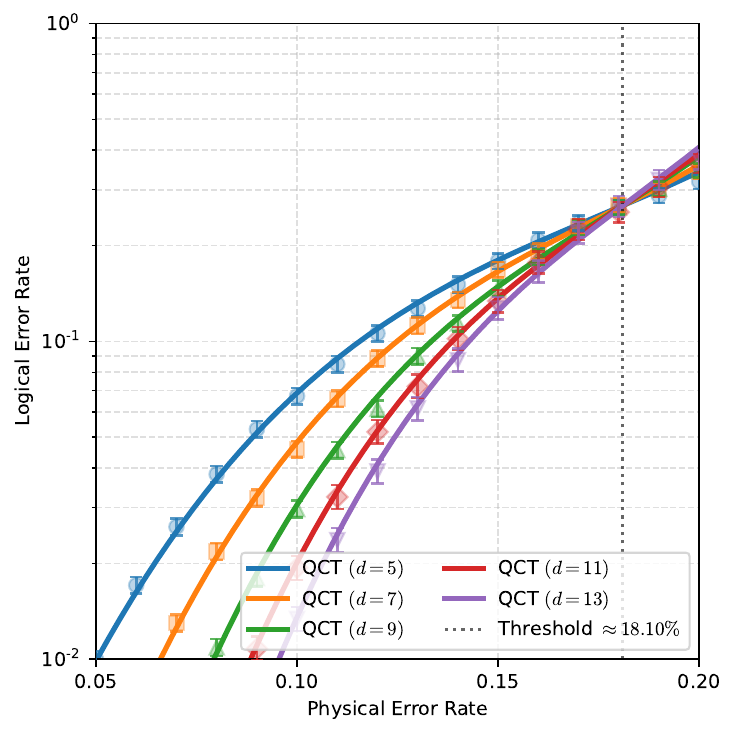}}
    \hfill
    \subfloat[]{\label{fig_BP_th}\includegraphics[width=\columnwidth]{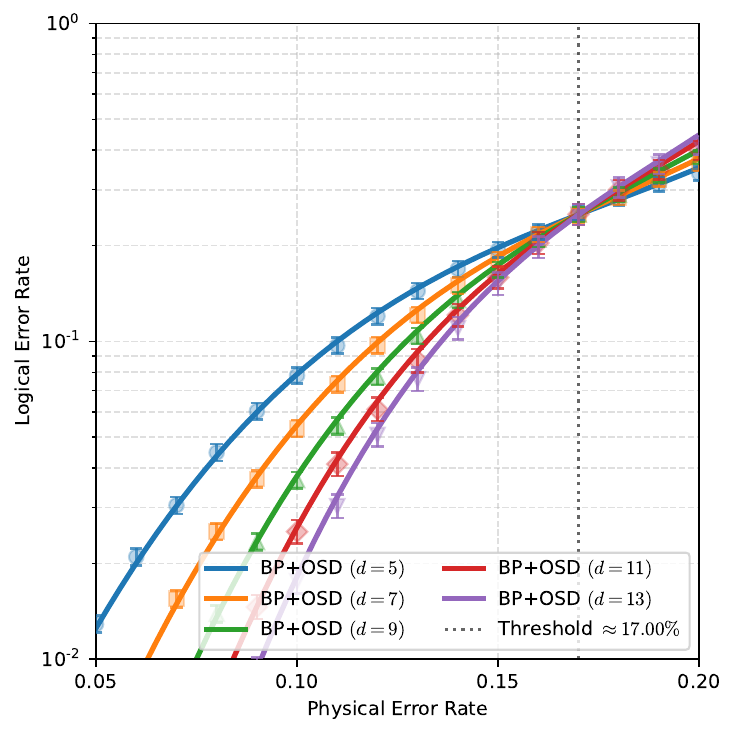}}
    \caption{\label{fig_th} 
    Threshold performance for surface codes with $d=5,7,9,11,13$.
    (a) QCT. (b) BP+OSD.
    At least $1000$ logical errors are collected for $d\in\{5,7,9\}$ and at least $500$ for $d\in\{11,13\}$. Error bars indicate 95\% Wilson confidence intervals.
    }
\end{figure}

\subsection{Model Size and Scalability}

Table~\ref{tab_params} summarizes the number of trainable parameters for
the evaluated code distances.
{For the code-capacity noise model}, the numbers of trainable parameters are
$1{,}609{,}076$, $1{,}613{,}732$, $1{,}619{,}940$, $1{,}627{,}700$, and
$1{,}637{,}012$ for $d=5,7,9,11,$ and $13$, respectively.
The parameter count increases linearly with the number of physical qubits
$n=d^2$, corresponding to $\mathcal{O}(n)=\mathcal{O}(d^2)$ scaling for
fixed $N$, $D_{\mathrm{e}}$, and $D_{\mathrm{m}}$.

For the circuit-level noise model, the original QCT transformer blocks are retained for within-round spatial processing, while additional temporal convolution blocks are introduced to capture temporal correlations across rounds.
The resulting models contain $2{,}004{,}690$, $2{,}020{,}338$, and $2{,}052{,}898$ trainable parameters for $d=3,5,$ and $7$, respectively,
showing only a small increase in model size over the evaluated code distances.

\subsection{Threshold Analysis}
To further quantify the decoder's performance, we compute the pseudo-threshold, defined as the physical error rate at which the LER equals that of un-coded qubits~\cite{aharonov1997fault}.
As presented in Table~\ref{tab_pseudo_th}, for a small-distance code ($d=3$), the pseudo-threshold of QCT is comparable to that of other neural network decoders and substantially higher than that of MWPM.
However, the performance gap widens with the code distance; for $d=5$ and $d=7$, QCT achieves the highest pseudo-threshold.
This result demonstrates that QCT not only offers stronger error-correction capability but also scales more effectively with increasing code size.

The code threshold is determined by identifying the crossing point of the LER curves for code distances ranging from $d=5$ to $d=13$. 
Fig.~\ref{fig_th} compares the LER performance and threshold of QCT with the BP+OSD decoder.
As shown in Fig.~\ref{fig_th}(a), QCT achieves a high threshold of $18.1\%$ {under the code-capacity noise model}, which closely approaches the theoretical bound of $18.9\%$~\cite{Bombin2012}.
This value outperforms both the BP+OSD decoder with a threshold of $17.00$\% as shown in Fig.~\ref{fig_th}(b), and the conventional MWPM decoder~\cite{Edmonds1965} with a threshold of $14.7$\%. 
The superior threshold of QCT indicates that the qubit-centric representation and structure-aware attention mechanism effectively capture the structure of the surface code.

\section{Conclusion}
\label{sec_conclusion}

In this work, we proposed QCT, a transformer-based QEC decoder that
organizes syndrome information around physical qubits through a
qubit-centric representation.
The embedding and merging layers construct unified physical-qubit tokens,
while the structure-aware mask incorporates the connectivity induced by
the stabilizer structure into the attention mechanism.
For circuit-level noise, we further extended QCT to repeated syndrome
extraction by preserving the qubit-centric spatial processing within each
round and interleaving temporal convolution blocks to capture temporal
correlations across rounds.
{Under the code-capacity noise model}, QCT achieves strong decoding performance across the evaluated surface-code distances.
In particular, QCT achieves a threshold of $18.1\%$, compared with
$17.00\%$ for BP+OSD and $14.7\%$ for MWPM.
The error suppression factor further shows stronger distance-dependent
logical error suppression for QCT than for the baseline decoders.
Under the circuit-level noise model, QCT also consistently achieves lower
logical error rates than standard MWPM, correlated MWPM, and BP+OSD across
the evaluated code distances and physical error rates.
The corresponding $\Lambda_{5/7}$ results further demonstrate strong
distance-dependent error suppression in the presence of circuit-level
faults.
Several directions remain for further investigation, including the extension
of QCT beyond surface codes to more general quantum low-density parity-check~(qLDPC) codes such as bivariate bicycle~(BB) codes.
Despite its strong performance, QCT introduces additional computational overhead, motivating more lightweight and hardware-efficient designs.
Furthermore, its robustness should also be examined under more hardware-relevant noise environments, including the SI1000 circuit-level noise model, crosstalk, leakage, and spatially nonuniform device noise.

\section{Acknowledgements}

This work was supported in part by the Basic Science Research
Program through the National Research Foundation of Korea (NRF) funded by the Ministry of Education under Grant RS-2026-25602443, and in part by Institute of Information \& communications Technology Planning \& Evaluation (IITP) grant funded by the Korea government (MSIT) (RS-2024-00399401, Development of Quantum-Safe Infrastructure Migration and Quantum Security Verification Technologies).

\bibliography{apssamp}

\end{document}